\documentclass[10pt,conference]{IEEEtran}
\usepackage[cmex10]{amsmath}
\usepackage{amssymb}
\usepackage[mathscr]{eucal}  
\usepackage{url}

\newtheorem{Proposition}{Proposition}
\newtheorem{Corollary}{Corollary}
\newtheorem{Theorem}{Theorem}
\newtheorem{Lemma}{Lemma}

\newtheorem{Definition}{Definition}
\newtheorem{Observation}{Observation}
\newtheorem{Remark}{Remark}
\newcommand{\eot}{\ $\square$} 
\newcommand{\ZU}{\ensuremath{\tilde{Z}}} 

\begin{document}


\title{On the Rate of Channel Polarization}
\author{\IEEEauthorblockN{Erdal Ar{\i}kan}
\IEEEauthorblockA{Department of Electrical-Electronics Engineering\\
Bilkent University\\
Ankara, TR-06800, Turkey\\
Email: arikan@ee.bilkent.edu.tr}\\
\and
\IEEEauthorblockN{Emre Telatar}
\IEEEauthorblockA{Information Theory Laboratory\\
Ecole Polytechnique F\' ed\' erale de Lausanne\\
CH-1015 Lausanne, Switzerland\\
Email: emre.telatar@epfl.ch}}
\maketitle

\begin{abstract} 
It is shown that for any binary-input discrete memoryless channel $W$ with symmetric capacity $I(W)$ and any rate $R <I(W)$, the probability of block decoding error for polar coding under successive cancellation decoding satisfies $P_e \le 2^{-N^\beta}$ for any $\beta<\frac12$ when the block-length $N$ is large enough.
\end{abstract}
\section{Introduction}\label{sec:Introduction}

Channel polarization is a method, introduced in \cite{ArikanPolarization}, for constructing a class of capacity-achieving codes, called {\sl polar codes}, on binary-input symmetric channels. 
Polar codes are of interest theoretically because they have a well-defined construction rule (that involves no trial-and-error) and are {\sl provably} capacity-achieving. 
The aim of this paper is to strengthen the results of \cite{ArikanPolarization} on the probability of block decoding error for polar codes. We begin by giving the notation and the general problem set-up.

Let $W:{\mathcal X}\to {\mathcal Y}$ be an arbitrary binary-input DMC (B-DMC) with input alphabet ${\mathcal X}=\{0,1\}$, output alphabet ${\mathcal Y}$, and transition probabilities $\{W(y|x):x\in{\mathcal X}, y\in {\mathcal Y}\}$. 
Let $I(W)$ denote the {\sl symmetric capacity} of $W$ defined as the mutual information (in bits) between the input and output terminals of $W$ when the input is chosen from the uniform distribution on ${\mathcal X}$.
This parameter takes values in $[0,1]$ and sets a limit on achievable rates across the channel $W$ using codes that employ the channel input letters with equal frequency.
Let $Z(W)= \sum_{y\in {\mathcal Y}} \sqrt{W(y|0)W(y|1)}$.
This parameter also takes values in $[0,1]$ and is an upper bound on the probability of ML decision error when the channel is used only once to transmit either a 0 or a 1. We will use $Z(W)$ as a measure of reliability.
 
The parameter $I(W)$ is of a more fundamental nature than $Z(W)$, however, $Z(W)$ will play a more central role in the following analysis since it is more readily tractable.
A useful pair of inequalities that relate these two parameters are
\begin{align}\label{ineq:IZ}
\begin{split}
I(W)^2 + Z(W)^2 & \le 1,\\
I(W) + Z(W) & \ge 1,
\end{split}
\end{align}
both proved in \cite{ArikanPolarization}.

\subsection{A channel transform}
Let ${\mathcal W}$ denote the class of all B-DMCs as defined above.
Consider a channel transform $W \mapsto (W^-,W^+)$ that maps ${\mathcal W}$ to ${\mathcal W}^2$.
Suppose the transform operates on an input channel $W:{\mathcal X} \to {\mathcal Y}$ to generate the channels $W^-:{\mathcal X} \to {\mathcal Y}^2$ and $W^{+}:{\mathcal X} \to {\mathcal Y}^2 \times {\mathcal X}$ with transition probabilities
\begin{align}
\begin{split}
W^{-}(y_1y_2|x_1) & = \sum_{x_2\in {\mathcal X}} \frac12 W(y_1|x_1\oplus x_2) W(y_2|x_2),\\
W^{+}(y_1y_2x_1|x_2) & = \frac12 W(y_1|x_1\oplus x_2) W(y_2|x_2),
\end{split}
\end{align}
where $\oplus$ denotes mod-2 addition.

Notice that in an actual implementation of this transform one needs two independent copies of $W$ to generate $W^-$ and $W^{+}$.
In that sense, the transform preserves symmetric capacity,
\begin{align}\label{eq:ConservationI}
I(W^-) + I(W^{+}) & = 2 I(W),
\end{align}
which is a direct consequence of the chain rule of mutual information.
As for the other parameter, we have
\begin{align}\label{eq:Zproperties}
\begin{split}
Z(W^{+})  & = Z(W)^2\\
Z(W) \le Z(W^{-}) & \le 2 Z(W)-Z(W)^2
\end{split}
\end{align}
whose proofs can be found in \cite{ArikanPolarization}.
Thus, the overall reliability is improved in the sense that 
\begin{align}\label{ineq:Zsuperadditivity}
Z(W^{-}) + Z(W^{+}) & \le 2 Z(W),
\end{align}
with $W^+$ more reliable than $W$ and $W^-$ less reliable than $W$. 

\subsection{Polarization process}\label{sec:polproc}

Let $(\Omega,{\cal F},P)$ be a probability space and suppose that $\{B_n:n=1,2,\ldots\}$ is a sequence of i.i.d. random variables defined on this space with
\begin{align}\label{eq:B_process}
P(B_1=0) & = P(B_1 = 1) = \frac{1}{2}.
\end{align}
For $n\ge 1$, let ${\mathcal F}_n$ be the $\sigma$-algebra generated by $(B_1,\ldots,B_n)$.
We may take ${\mathcal F} = \cup_{n=1}^\infty {\mathcal F}_n$. 

Fix a channel $W\in {\mathcal W}$. Define a random sequence of channels $\{W_n\in {\mathcal W}: n\ge 0\}$ that starts at $W_0 = W$, and at time $n\ge 1$ sets 
\begin{align}\label{eq:DefWn}
W_{n} & = \begin{cases} W_{n-1}^-  & \text{if $B_{n}=1$},\\
  W_{n-1}^{+} & \text{if $B_{n}=0$},
  \end{cases}
\end{align}
where the channels on the right side are defined by the transform $W_{n-1} \mapsto (W_{n-1}^-,W_{n-1}^{+})$.
Define two random processes $\{I_n:n=0,1,\ldots\}$ and $\{Z_n:n=0,1,\ldots\}$
by setting $I_n:= I(W_n)$ and $Z_n:= Z(W_n)$.

\begin{Observation}\mbox{}
\begin{enumerate}
\item[(i)] $\{(I_n,{\mathcal F}_n)\}$ is a bounded martingale on $[0,1]$ and converges a.s. to a r.v. $I_\infty$.
\item[(ii)] $\{(Z_n,{\mathcal F}_n)\}$ is a bounded supermartingale on $[0,1]$ and converges a.s. to a r.v. $Z_\infty$.
\end{enumerate}
\end{Observation}
The martingale and supermartingale properties follow from \eqref{eq:ConservationI}, \eqref{ineq:Zsuperadditivity}, and the convergence properties from general results on bounded martingales. It was shown in \cite{ArikanPolarization} and we will show in the sequel that the limit random variables $I_\infty$ and $Z_\infty$ are a.s. 0-1 valued. It then follows that $I_\infty + Z_\infty = 1$ in view of \eqref{ineq:IZ}.
Since $E[I_\infty] = I_0 = I(W)$, we have $P(I_\infty = 1) = I(W)$ and $P(I_\infty = 0) = 1-I(W)$.  Consequently $P(Z_\infty=0)=I(W)$ and $P(Z_\infty=1)=1-I(W)$.
Thus the sequence of channels $\{W_n\}$ \emph{polarizes} with probability one: they become perfect with probability $I(W)$, useless with probability $1-I(W)$.

\subsection{Polar coding}
Channel polarization was used in \cite{ArikanPolarization} to develop a channel coding scheme called \emph{polar coding}.
Polar codes are a class of block codes with block-lengths constrained to $N=2^n$, $n\ge 0$.
These codes can be encoded in complexity $O(N\log N)$ and decoded by a successive cancellation decoder also in complexity $O(N\log N)$.
These complexity bounds hold uniformly for all rates $R\in [0,1]$, although for $R>I(W)$, they have no practical relevance.

To state the results precisely, let $P_e(N,R)$ denote the best achievable block error probability under successive cancellation decoding for polar coding with block length $N$ and rate $R$.
It was shown in \cite{ArikanPolarization} that for any given channel $W\in {\mathcal W}$, any $n$, and any $\gamma\in[0,1]$, there exists a polar code with block-length $N=2^n$, whose rate $R$ and probability of block error under successive cancellation decoding $P_e$ satisfy
\begin{align}
\label{eq:ParametricR} R & \ge P(Z_n \le \gamma) \\
\label{eq:ParametricP} P_e & \le N \gamma.
\end{align}
The main result of \cite{ArikanPolarization} in this regard was to show that for any $R<I(W)$ the relation~\eqref{eq:ParametricR} can be satisfied for large $N$ by taking the parameter $\gamma$ as a function $\gamma(N,R) = o(N^{-\frac{5}{4}})$.  
This enabled \cite{ArikanPolarization} to conclude from~\eqref{eq:ParametricP} that $P_e(N,R) = o(N^{-\frac14})$ for any fixed $R <I(W)$.

\subsection{Summary of results}

In this paper we improve the results of \cite{ArikanPolarization} by proving the following 
\begin{Theorem}\label{Thm:PolarCode}
Let $W$ be any B-DMC with $I(W) >0$. Let $R < I(W)$ and $\beta < \frac12$ be fixed. Then, for $N=2^n$, $n\ge 0$, the best achievable block error probability for polar coding under successive cancellation decoding at block length $N$ and rate $R$ satisfies
\begin{align}\label{eq:PeUB}
P_e(N,R) & = o(2^{-N^\beta}).
\end{align}  
\eot
\end{Theorem}

\begin{Remark} 
The bound \eqref{eq:PeUB} depends only on whether $R<I(W)$, but otherwise is not sensitive to $R$. 
Determining sharper asymptotic results on $P_e(N,R)$ that display a more refined dependence on $R$ remains a challenging open problem.
\end{Remark}

This result will follow from \eqref{eq:ParametricR} and \eqref{eq:ParametricP} as a corollary to the first half of the following
\begin{Theorem}\label{Thm:Polarization}
Let $W$ be any B-DMC. 
For any fixed $\beta < \frac12$, 
\begin{align}\label{eq:gamma1}
\liminf_{n\to \infty} P(Z_n \le 2^{-N^\beta}) & = I(W).
\end{align}
Conversely, if $I(W) <1$, then for any fixed $\beta > \frac12$, 
\begin{align}\label{eq:gamma2}
\liminf_{n\to \infty} P(Z_n \ge 2^{-N^\beta}) & = 1.
\end{align}
\eot
\end{Theorem}


The rest of this paper is devoted to proving Theorem~\ref{Thm:Polarization}.
The analysis will be carried out using the supermartingale $\{Z_n\}$. 
Section~\ref{sec:Zprocesses} abstracts out the general properties of this supermartingale so as to carry out the analysis is a more general framework unencumbered by the details of the original information-theoretic context. 
Theorem~\ref{Thm:Polarization} is restated in Section~\ref{sec:Zprocesses} in a general setting and proved in the sections that follow.
In Section~\ref{sec:Conclusions}, we state some open problems.

\section{Problem restatement}\label{sec:Zprocesses}

Let the probability space $(\Omega,{\cal F},P)$, the Bernoulli sequence $\{B_n:n=1,2,\ldots\}$, and the $\sigma$-algebras $\{{\mathcal F}_n\}$ be defined in Section~\ref{sec:polproc} above.
We define the following class of random processes on $(\Omega,{\cal F},P)$.
\begin{Definition}\label{def:Zprocess}
For each $z_0 \in (0,1)$, define ${\mathcal Z}_{z_0}$ as the class of random processes $\{Z_n: n=0,1,\ldots\}$ such that the process begins at $Z_0 = z_0$, $Z_n$ is measureable with respect to ${\mathcal F}_n$, and follows trajectories satisfying
\begin{align}\label{DefinitionZ}
Z_{n+1} & = Z_{n}^2 \quad \text{if $B_{n+1}=1$},\\
Z_{n+1} & \in [Z_{n}, 2Z_{n} - Z_{n}^2] \quad \text{if $B_{n+1}=0$},
\end{align}
for $n\ge 0$.
Let ${\mathcal Z}:= \cup_{z_0 \in (0,1)}{\mathcal Z}_{z_0}$.
\end{Definition}

The class ${\mathcal Z}$ contains the processes $\{Z_n\}$ that were defined in Section~\ref{sec:Introduction} for all non-trivial channels $W\in {\mathcal W}$ for which $0< Z(W) <1$.
The cases $z_0=0$ and $z_0=1$ are excluded from the definition since these lead to trivial processes which only complicate the statement of the results.
Notice that the definition of ${\mathcal Z}$ makes no reference to the information-theoretic origin of the problem, making the rest of the discussion fully self-contained.

\begin{Observation} \label{Zproperties}
For any $\{Z_n\}\in {\mathcal Z}$, the following hold.
\begin{enumerate}
\item[(i)] $Z_{n} \in (0,1)$ for all $n\ge 0$.
\item[(ii)] $\{(Z_n,{\mathcal F}_n)\}$ is a bounded supermartingale.
\item[(iii)] $\{Z_n\}$ converges a.s. and in ${\mathcal L}^1$ to a random variable $Z_\infty$, which is 0-1 valued a.s.
\end{enumerate}
\end{Observation}

The first two observations are obvious.
That $\{Z_n\}$ converges a.s. and in ${\mathcal L}^1$ is by general theorems on bounded supermartingales. 
Convergence in ${\mathcal L}^1$ implies that $E[|Z_{n+1}-Z_{n}|] \to 0$.
But, $E[|Z_{n+1}-Z_n|] \ge (1/2)E[Z_n-Z_n^2]\ge 0$, which implies $E[Z_n(1-Z_n)] \to 0$ and $E[Z_\infty(1-Z_\infty)] = 0$. 
Thus $Z_\infty$ equals 0 or 1 a.s.

We will prove Theorem~\ref{Thm:Polarization} by proving the following equivalent
\begin{Theorem}\label{Thm:MainBound}
For any $\{Z_n\} \in {\mathcal Z}$ and $\beta < \frac12$, we have 
\begin{align}\label{eq:Direct}
\liminf_{n\to \infty} P(Z_n \le 2^{-2^{\beta n}}) & \ge P(Z_\infty = 0);
\end{align}
conversely, for $\beta > \frac{1}{2}$, 
\begin{align}\label{eq:Converse}
\liminf_{n\to \infty} P(Z_n \ge 2^{-2^{\beta n}} ) & = 1.
\end{align}
\end{Theorem} 

The proof of the converse part \eqref{eq:Converse} will be given in the next section. The direct part \eqref{eq:Direct} will be proved in Section~\ref{sec:DirectPart}.

\section{Proof of the converse part}\label{sec:Converse}

Fix a process $\{Z_n\}\in {\mathcal Z}$.  Fix $\beta>1/2$ and put $\delta_n(\beta):= 2^{-2^{\beta n}}$. 

Let $\{\ZU_i\}$ be defined as the random process
\begin{align*}
\ZU_0=Z_0,\quad
\ZU_{i+1} =\begin{cases} \ZU_i^2 & \text{if $B_{i+1}=1$}\\
\ZU_i & \text{if $B_{i+1}=0$}
\end{cases}\quad i\geq0.
\end{align*}
A comparison with \eqref{Zrecursion} shows that $\{\ZU_i\}$ is dominated by $\{Z_i\}\in {\mathcal Z}$ and thus,
\begin{align}
P(Z_n \ge \delta_n) \ge  P(\ZU_n \ge \delta_n)
\end{align}
Notice that 
\begin{align}
\ZU_{n} = Z_0^{\bigl(2^L\bigr)}
\end{align}
with $L=\sum_{i=1}^n B_i$.  Thus,
\begin{align}\label{eq:Binom}
P(\ZU_n \ge \delta_n) = P(L +\log_2\log_2(1/Z_0) \le n\beta).
\end{align}
As $\beta >\frac{1}{2}$ and $Z_0>0$, by the law of large numbers, this probability goes to 1 as $n$ increases, yielding~\eqref{eq:Converse}.

\section{Proof of the direct part}\label{sec:DirectPart}

\begin{Definition}
Given a process $\{Z_n\}\in {\mathcal Z}$ and a sequence of reals $\{f_n\}\subset [0,1]$ convergent to 0, we will say that $\{f_n\}$ is {\sl asymptotically dominating} (a.d.) for $\{Z_n\}$ and write $Z_n \prec f_n$ to mean that
\begin{align*}
\liminf_{n\to \infty} P(Z_n \le f_{n}) & \ge P(Z_\infty = 0).
\end{align*}
We will say that $\{f_n\}$ is {\sl universally dominating} (u.d.) for $\{Z_n\}$ if, for any fixed $k\ge 0$, the sequence $\{f_{n+k}\}$ is a.d. for $\{Z_n\}$.
\end{Definition}

In this notation, the direct part of Theorem~\ref{Thm:MainBound} claims that, for $\beta < \frac12$, the sequence $2^{-2^{\beta n}}$ is a.d. for every process in ${\mathcal Z}$.
We will prove this claim in several steps. 
First, we define in Section~\ref{subsec:Extremal} a subclass of processes in ${\cal Z}$ called \emph{extremal processes}.
Next, we show in Section~\ref{subsec:Reduction} that a sequence $\{f_n\}$ is a.d. for the class ${\mathcal Z}$ if it is u.d. for the subclass of extremal processes. 
In Section~\ref{subsec:Exponential}, we show that $\{\rho^n\}$ with $\rho \in (\frac34,1)$ is a.d. for every extremal process. In Section~\ref{subsec:Bootstrap}, we use this result to show that, for any fixed $\beta <\frac12$, the sequence $\{2^{-2^{n\beta}}\}$ is u.d. for extremal processes. 

\subsection{Extremal processes}\label{subsec:Extremal}

\begin{Definition}
A process $\{Z_n\}\in {\mathcal Z}$ is called {\sl extremal} if 
\begin{align}\label{Zrecursion}
Z_{n+1}  & = \begin{cases} Z_n^2 & \text{if $B_{n+1} = 1$},\\
2Z_n -Z_n^2 & \text{if $B_{n+1} = 0$}.
\end{cases}
\end{align}
The extremal process in ${\mathcal Z}_{z_0}$ will be denoted by the notation $\{Z^{(z_0)}_{n}\}$ when we need to refer to it explicitly.
\end{Definition}

Note that the recursion for an extremal process can be written alternatively as 
\begin{align}
Z_{n+1}  & = Z_n^2  \qquad \text{if $B_{n+1} = 1$}\\
(1-Z_{n+1}) & = (1-Z_n)^2 \qquad \text{if $B_{n+1} = 0$}.
\end{align}
and also as
\begin{align}
Z_{n+1} = Z_n + X_n Z_n(1-Z_n), \quad n \ge 0
\end{align}
where $X_n = (1-2B_n)$ is a $\pm 1$-valued random process.
These forms emphasize the symmetric nature of the extremal process.

We state some properties of extremal processes that follow immediately from Observation~\ref{Zproperties}.
\begin{Observation} \label{ZpropertiesExtemal} For $\{ Z_n\}$ any extremal process, in addition to Observation~\ref{Zproperties}, we have
\begin{enumerate}
\item[(i)] $\{Z_n\}$ is a Markov process.
\item[(ii)] $\{Z_n\}$ is a bounded martingale.
\item[(iii)] $P(Z_\infty=0)=1-Z_0,\quad P(Z_\infty=1) = Z_0$.
\end{enumerate}
\end{Observation}

The term \emph{extremal} is justified by the following
\begin{Observation}\label{Zproperties3}
\mbox{}
\begin{enumerate}
\item[(i)] Every process $\{Z_n\} \in {\mathcal Z}_{z_0}$ is dominated by $\{Z^{(z_0)}_{n}\}$ on a sample function basis, i.e., $Z_n\leq Z^{(z_0)}_n$.
\item[(ii)] The extremal process $\{Z_n^{(\alpha)}\}$ is dominated by $\{Z_n^{(\beta)}\}$ on a sample function basis for all $0 < \alpha \le \beta < 1$.
\end{enumerate}
\end{Observation}

\subsection{A reduction argument}\label{subsec:Reduction}

\begin{Proposition}
If $\{f_n\}$ is a u.d. sequence over the class of extremal processes in ${\mathcal Z}$, then $\{f_n\}$ is a.d. over the class ${\mathcal Z}$. 
\end{Proposition}

\begin{IEEEproof}
Fix a process $\{Z_n\}$ in ${\mathcal Z}$ and a sequence $\{f_n\}$ that is u.d. over the class of extremal processes.
For any $k\ge 0$, $n\ge 0$, and $\delta \in (0,1)$, we trivially have
\begin{align}\label{eq:1}
P(Z_{k+n} \le f_{k+n}) & \ge P(Z_{k+n} \le f_{k+n} \mid  Z_k \le \delta) \,P(Z_k \le \delta).
\end{align}
Combining the observations
\begin{align*}
P(Z_{k+n} \le f_{k+n} \mid  Z_k \le \delta) \;\ge \;P(Z_n^{(\delta)} \le f_{k+n})
\end{align*}
and
\begin{align*}
\liminf_{n\to \infty} P(Z_n^{(\delta)} \le f_{n+k}) \ge (1-\delta)
\end{align*}
with \eqref{eq:1}, 
we see that for any fixed $k\ge 0$
\begin{align*}
\liminf_{n\to \infty} P(Z_{n} \le f_{n}) & = \liminf_{n\to \infty} P(Z_{n+k} \le f_{n+k})\\
&  \ge (1-\delta) P(Z_k \le \delta).
\end{align*}
Since this is true for all $k$, we obtain
\begin{align*}
\liminf_{n\to \infty} P(Z_{n} \le f_{n}) & \ge  (1-\delta) \,\liminf_{k\to \infty} P(Z_k \le \delta)\\
& \ge (1-\delta) \,P(\liminf_{k\to \infty} Z_k \le \delta)\\
& = (1-\delta)\,P(Z_\infty = 0)
\end{align*}
where the second line follows by Fatou's lemma and the third by the a.s. convergence of $\{Z_k\}$ to the 0-1 valued $Z_\infty$.
Letting $\delta \to 0^+$, we obtain
\begin{align*}
\liminf_{n\to \infty} P(Z_{n} \le f_{n}) & \ge P(Z_\infty = 0),
\end{align*}
which completes the proof.
\end{IEEEproof}

\subsection{An asymptotically dominating sequence}\label{subsec:Exponential}

\begin{Proposition}\label{prop:Exponential} For any $\rho \in(\frac34,1)$, the sequence $\{\rho^n\}$ is a.d. over the class of extremal processes.
\end{Proposition}

To prove this statement, let us fix $\{Z_n\}$ as an extremal process in ${\mathcal Z}$ with $Z_0=z_0$ for some $z_0\in (0,1)$. 

Let $Q_n:= Z_n(1-Z_n)$. Then $Q_n \in (0,\frac{1}{4}]$ and
\begin{align}
Q_{n+1} & = 
\begin{cases} 
Z_n^2 (1-Z_n^2) & \text{if $B_{n+1}= 1$}\\
(2Z_n-Z_n^2)(1-2Z_n+Z_n^2) & \text{if $B_{n+1}= 0$}
\end{cases}\nonumber\\
& = Q_n \cdot
\begin{cases} 
Z_n (1+Z_n) & \text{if $B_{n+1}= 1$}\\
(1-Z_n) (2-Z_n) & \text{if $B_{n+1}= 0$}.
\end{cases}\label{Qrecursion}
\end{align}

\begin{Lemma}[\cite{Hajek}] $E[Q_n^{1/2}] \le \frac{1}{2} \left(\frac{3}{4}\right)^{n/2}$.
\end{Lemma}
\begin{IEEEproof}
Note that $\sqrt{z(1+z)} + \sqrt{(1-z)(2-z)} \le \sqrt{3}$ when $z\in [0,1]$.
So, by \eqref{Qrecursion}, $E[Q_{n+1}^{1/2}\,|\, Q_n] \le Q_n^{1/2} \left(\frac{3}{4}\right)^{1/2}$.
Thus $E[Q_n^{1/2}] \le E[Q_0^{1/2}] \left(\frac{3}{4}\right)^{n/2} \le \frac{1}{2}\left(\frac{3}{4}\right)^{n/2}$.
\end{IEEEproof}

By Markov's inequality, we obtain
\begin{Corollary}\label{cor:Q} $P(Q_n \ge \rho^n) \le \frac{1}{2} \left(\frac{3}{4\rho}\right)^{n/2}$ for $\rho >0$.
\end{Corollary}

We now turn this into a bound on $Z_n$.
\begin{Lemma}\label{lem:rho} Let $f_n(\rho):= \frac{1-\sqrt{1-4\rho^n}}{2}$ if $1-4\rho^n>0$, $f_n(\rho):=1$ otherwise.
Then, $Z_n \prec f_n(\rho)$ for all $\rho\in (\frac{3}{4},1)$.
\end{Lemma}
\begin{IEEEproof}
Fix $\rho\in (\frac{3}{4},1)$ and let $f_n=f_n(\rho)$.
Note that for $n$ large enough so that $1-4\rho^n >0$, we have
\begin{align} 
\{Q_n\le \rho^n\} & = \{Z_n\le f_n\} \cup \{Z_n\ge 1-f_n\}
\end{align}
where the sets on the right side are disjoint. So, for $n$ large enough
\begin{align}
P(Q_n \le \rho^n)  & = P(Z_n \le f_n) + P(Z_n \ge 1-f_n)
\end{align}
which gives
\begin{align}
\liminf_{n\to \infty} P(Q_n \le \rho^n)  & \le \liminf_{n\to \infty}P(Z_n \le f_n) \nonumber\\ & + \limsup_{n\to \infty}P(Z_n \ge 1-f_n).
\end{align}
Since $\rho \ge \frac34$, the left side of the above equation equals 1 by Corollary~\ref{cor:Q}.
Since $f_n$ is monotonically decreasing,
\begin{align}
\limsup_{n\to \infty}P(Z_n \ge 1-f_n) & \le \limsup_{n\to \infty}P(Z_n \ge 1-f_k)
 \end{align}
 for any $k\ge 1$.
 But $\limsup_{n\to \infty}P(Z_n \ge 1-f_k) = z_0$.
 Thus, 
\begin{align}
\liminf_{n\to \infty}P(Z_n \le f_n)  & \ge 1 -z_0,
\end{align}
which means that $Z_n\prec f_n$, as claimed. 
\end{IEEEproof}

The proof of Proposition~\ref{prop:Exponential} will be complete if we show that for every $\rho\in (\frac{3}{4},1)$, there exists $\tilde{\rho} \in (\frac{3}{4},1)$, such that $f_n(\tilde{\rho})\le \rho^{n}$ for all $n$ large enough.
It is easy to see that this is true for any $\frac 34 <\tilde{\rho} < \rho$.
\
\subsection{A bootstrapping argument}\label{subsec:Bootstrap}

We now strengthen the result of the previous subsection and complete the proof of the direct part of Theorem~\ref{Thm:MainBound}.

\begin{Proposition}\label{prop:UD}
For any $\beta <\frac12$, the sequence $\{2^{-2^{n\beta}}\}$ is u.d. over the class of extremal processes.
\end{Proposition}

\begin{IEEEproof}
Fix $\beta <\frac12$. 
First note that, for any fixed $k\ge 0$, asymptotically in $n$, we have $2^{-2^{(n+k)\beta}} = \Theta(2^{-2^{n\beta}})$ (using standard Landau notation). Hence, it suffices to prove that $\{2^{-2^{n\beta}}\}$ is an a.d. sequence.

Fix $\{Z_n\}$ as an extremal process. We wish to prove that $Z_n \prec 2^{-2^{\beta n}}$.
Consider a second process $\{\tilde{Z}_i\}$ defined by fixing an $n\ge 1$ and an $m\in \{0,\ldots,n\}$ and setting
\begin{align*}
\tilde{Z}_i & =Z_i,\quad i=0,\ldots,m,\\
\tilde{Z}_{i+1} & =\begin{cases} \tilde{Z}_i^2 & \text{if $B_{i+1}=1$}\\
2\tilde{Z}_i & \text{if $B_{i+1}=0$}
\end{cases},
\quad i\ge m.
\end{align*}
A comparison with \eqref{Zrecursion} shows that $Z_i\le \tilde{Z}_i$ for all $i\ge 1$.

Fix $a_n= \sqrt{n}$, and partition the set $\{m,\ldots,n-1\}$ into $k= (n-m)/a_n$ consecutive intervals $J_1,\ldots,J_k$ of size $a_n$, i.e., $J_j =\{m+(j-1)a_n,\ldots,m+ja_n-1\}$.
Let $E_j$ be the event that $\sum_{i\in J_j} B_i < a_n\beta$.
Observe that 
\begin{align}
P(E_j) & \le 2^{-a_n[1-{\mathcal H}(\beta)]}
\end{align}
where ${\mathcal H}(\beta) = -\beta \log_2(\beta) -(1-\beta) \log_2(1-\beta)$ is the binary entropy function.
Thus the event $G:= \cap_j E_j^c$ has probability at least $1-k2^{-a_n[1-{\mathcal H}(\beta)]}$.
Conditional on $G$, during every interval $J_j$ the value of $\tilde Z$ is squared at least $a_n\beta$ times and doubled at most $a_n(1-\beta)$ times; hence, we have
\begin{align*}
\log_2 \tilde{Z}_{m+(j+1)a_n} \le 2^{a_n \beta} \Bigl[\log_2 \tilde{Z}_{m+ja_n} + a_n(1-\beta)\Bigr] 
\end{align*}
and so 
\begin{align*}
\log_2 Z_n & \le \log_2 \tilde{Z}_n \\
& \le 2^{(n-m)\beta} \log_2 Z_m +  a_n (1-\beta) \sum_{j=1}^k 2^{j a_n \beta}\\
& \le 2^{(n-m)\beta} \log_2 Z_m +  a_n (1-\beta) 2^{(n-m)\beta} \bigl(1- 2^{-a_n \beta}\bigr)^{-1}\\
& \le 2^{ (n-m)\beta} \bigl[\log_2 Z_m + a_n\bigr] \quad \text{for $n$ large enough.}
\end{align*}
Lastly, fix $m= n^{3/4}$, $\rho = 7/8$. Conditional on $\tilde G=\bigl\{Z_m \le \left(\frac{7}{8}\right)^m\bigr\} \cap G$ and for $n$ large enough, we have
$\log_2 Z_m \le -n^{3/4} \log_2(8/7)$; hence, 
\begin{align*}
\log_2 Z_n & \le 2^{(n-m)\beta}[-n^{3/4}\log_2(8/7)+n^{1/2}] \le -2^{n\beta} \,o(1)
\end{align*}
Noting that the probability of $G$ approaches $1$,
we see by Lemma~\ref{lem:rho} that the probability of $\tilde G$
approaches $1-z_0$.
This establishes that $Z_n \prec 2^{-2^{\beta n}}$ for any fixed $\beta <1/2$.
\end{IEEEproof}

\section{Open problems}\label{sec:Conclusions}

Broadly stated, we have been interested in the asymptotic behavior of the cumulative probabilities $P(Z_n\le z)$ for processes $\{Z_n\}$ derived from a channel polarization problem.
The ultimate result in this regard would be to determine explicitly a function $E(n,R)$ such that, for any $R\in [0,1]$,
\begin{align}
\liminf_{n\to \infty} P(Z_n \le 2^{-2^{E(n,R)}}) & = R.
\end{align}
Theorem~\ref{Thm:MainBound} gives only some partial characterization of $\frac{E(n,R)}{n}$ for large $n$.

The information-theoretic problem considered in this paper can be generalized in two main directions.
First, one may consider the transform $W\mapsto (W^-,W^{+})$ for channels with input alphabets ${\cal X}=\{1,\ldots,q\}$ for arbitrary $q\ge 2$.
In this generalization, the mod-2 addition operation $\oplus$ may be replaced with addition mod-$q$, or even with an arbitrary group operation on ${\cal X}$. 
The process $\{I_n\}$ can be defined as before, the conservation law \eqref{eq:ConservationI} still holds, and $\{I_n\}$ is a bounded martingale, which must converge a.s.
An initial open problem for this case is to prove that channel polarization takes place, i.e., that $\{I_n\}$ converges a.s. to the set $\{0,\log_2 q\}$. Conditional on the validity of channel polarization, a subsequent goal would be to determine the rate of polarization.

Note that for $q\ge 3$, the auxiliary random process $\{Z_n\}$ can be defined only after giving a new definition for the channel parameter $Z(W)$. A natural definition is $Z(W) = \sum_{x\neq x^\prime} \frac{1}{q(q-1)} \sum_{y} \sqrt{W(y|x)W(y|x^\prime)}$. Unfortunately, the relations \eqref{eq:Zproperties} do not hold for this definition, and the process $\{Z_n\}$ does not appear likely to facilitate the analysis for $q\ge 3$.

A second direction for generalization of the problem is to consider more general channel transformations that preserve mutual information. For example, a ternary operation $W\mapsto (W^\prime,W^{\prime\prime},W^{\prime\prime\prime})$ may be considered such that  $I(W^\prime)+I(W^{\prime\prime})+I(W^{\prime\prime\prime})=3I(W)$. 
The random sequence of channels $\{W_n\}$ can be defined using a ternary fair coin, which ensures that $\{I_n\}$ is a bounded martingale. A major open problem in this general setting is to determine necessary and sufficient conditions on the channel transform to ensure channel polarization. 

\section*{Acknowledgment}
The authors wish to thank anonymous reviewers for their valuable comments and corrections on an earlier version of this manuscript. 
This work was supported in part by The Scientific and
Technological Research Council of Turkey (T\"UB\.ITAK) under contracts no.
105E065 and 107E216.

\end{document}